\documentstyle[aps]{revtex}
\font\eightrm=cmr8
\def\bxk{{\scriptstyle({\eightrm x} \scriptstyle)}}
\def\bxpk{{\scriptstyle({\eightrm x}^{\eightrm\prime} \scriptstyle)}}
\def\bxdpk{{\scriptstyle({\eightrm x}^{\eightrm\prime\prime} \scriptstyle)}}
\def\lxr{{\scriptstyle[{\eightrm x} \scriptstyle]}}
\def\lxpr{{\scriptstyle[{\eightrm x}^{\eightrm\prime} \scriptstyle]}}
\def\bzk{{\scriptstyle({\eightrm z} \scriptstyle)}}
\def\bt0zk{{\scriptstyle({\eightrm t}_{\eightrm 0}, {\eightrm{\bf z}})}}
\def\mmbox#1#2{\vcenter{\hrule \hbox{\vrule height#2in
                \kern#1in \vrule} \hrule}}
\def\dAlemb{\mmbox{.09}{.09}}

\begin{document}
\draft
\begin{flushright} TOKAI-HEP/TH-0005 \end{flushright}
\vspace{0.5cm}
\centerline{\bf Classical theory of canonical QCD on a space-like hypersurface}
\vspace{1.0cm}
\centerline{Hiroshi Ozaki\footnote{Email address: ozaki@keyaki.cc.u-tokai.ac.jp}}
\vspace{0.5cm}
\centerline{\small {\it Department of Physics, Tokai University, 1117 Kitakaname, Hiratsuka 259-1292, Japan}}
\vspace{1.0cm}
\begin{abstract}
The canonical formalism in classical theory of QCD is constructed on a space-like hypersurface.
The Poisson bracket on the space-like hypersurface is defined and it plays
an important role to describe every algebraic relation in the canonical formalism
into Lorentz covariant form.
Surface integrals are introduced as alternatives of field equations for quarks, gluons, and
Faddeev-Popov ghosts. It is shown that deformations of the space-like hypersurface 
for surface integrals are generated by
the interaction term of QCD Hamiltonian density.
By converting the Poisson bracket on the space-like hypersurface to four-dimensional commutator, 
we can pass over to QCD in the Heisenberg picture without spoiling the explicit Lorentz covariance.
\end{abstract}
\pacs{ }

One of the problems in the ordinary canonical method of quantization in quantum chromodynamics (QCD)
is in the classical procedure blinding the Lorentz covariance.
The definition of the canonical conjugate momentum depends on the frame of reference.
The Hamiltonian is linked to a definite choice of the time.
Further, the commutation relation of two field variables is defined at the same time. 
From the relativistic point of view, the canonical formalism 
on the basis of a particular frame of reference
spoils the explicit Lorentz covariance in the classical description of QCD 
far from field quantization. 
For this reason the canonical method of quantization in QCD
have been regarded to be inferior to path-integral method.
In the path-integral method Lorentz covariance is manifest from the outset.
This is an impressive method to construct the S-matrix in QCD with covariant manner. 
Although the canonical method is superior to the path-integral method in presenting clearly a particle 
interpretation to quarks, gluons, and Faddeev-Popov ghosts, 
the lack of explicit Lorentz covariance turns our interest to the path-integral method.
This paper presents a canonical formalism in the classical theory of QCD
with coherently Lorentz covariant form. 
The main subject is to construct the canonical formalism on a space-like hypersurface.
It is shown that
the canonical method of quatization in QCD should be applied on the space-like hypersurface
by converting classical fields to quantized field operators in the Heisenberg picture.
The generalization of our formulation will be discussed in separate papers.    
 
The total QCD Lagrangian density with $\xi=1$ gauge is taken as 
\begin{eqnarray}
{\cal L}\lxr  &=&\bar \psi\bxk \left( {i\gamma ^\mu \partial _\mu 
  -m+g\gamma ^\mu \lambda ^iW_\mu ^i \bxk} \right)\psi\bxk
  -{1 \over 4}G_{\mu \nu }^i \bxk G^{i\mu \nu }\bxk
  -{1 \over 2}\left( {\partial ^\mu W_\mu ^i \bxk} \right)^2 \nonumber \\
  &+&{1 \over 2}\partial _\mu \left[ {W_\nu ^i \bxk\left( {g^{\mu \nu }\partial _\rho W^{i\rho }\bxk
  -\partial _\nu W^{i\mu } \bxk} \right)} \right]
  +\left( {\partial ^\mu \bar c^a \bxk} \right)D_\mu ^{ab}c^b \bxk, \label{eq:LAG}
\end{eqnarray}
with
\begin{eqnarray}
  G_{\mu \nu }^i \bxk&=&\partial _\mu W_\nu ^i \bxk-\partial _\nu W_\mu ^i \bxk
  +g f^{ijk}W_\mu^j \bxk W_\nu ^k \bxk,\nonumber \\
  D_\mu ^{ab}&=&\delta ^{ab}\partial _\mu +g f^{acb}W_\mu ^c \bxk, \nonumber
\end{eqnarray}
where $\psi$ is the $SU_c (3)$ triplet quark, $W_{\mu}^i$ are gluon fields, $c^a$ and $\bar c^a$ are
Faddeev-Popov ghosts and anti-ghosts, $\lambda^i$ are the Gell-Mann matrices, 
and $f^{ijk}$ are the structure constants of $SU_c (3)$.
With the Lagrangian density (\ref{eq:LAG}), the principle of the least action gives field equations
\begin{eqnarray}
& &\left( {i\gamma ^\mu \partial _\mu -m+g\gamma ^\mu \lambda ^iW_\mu ^i \bxk} \right)\psi\bxk =0,\label{eq:EL1}\\
& &\partial _\mu \left[ {\partial ^\mu W^{i\rho }\bxk
  +g f^{ijk}W^{j\mu } \bxk W^{k\rho } \bxk} \right]
  -g f^{nik}W_\nu ^k \bxk G^{n\rho \nu }\bxk \nonumber\\
& &\qquad\qquad +g f^{iab} [ {\partial ^\rho \bar c^a \bxk} ] c^b \bxk
  -g\bar \psi \bxk\gamma ^\rho \lambda ^i\psi \bxk=0,\label{eq:EL2}\\
& &  \partial ^\mu D_\mu ^{ab}c^b \bxk=0,\label{eq:EL3}\\
& &  D^{ab\mu }\partial _\mu \bar c^b \bxk=0.\label{eq:EL4}
\end{eqnarray}

In QCD all field variables appeared in the Lagrangian density are field operators. However,
we classicize the gluon field operators into Grassmann even wave functions, 
and the quark field and Faddeev-Popov ghost operators into Grassmann odd wave functions 
to look into only the classical aspect of QCD.

The canonical formalism needs the canonical variables. We will adopt
every classical field and its conjugate momentum on a space-like hypersurface
as the canonical variables.
Let us choose conjugate momenta on the space-like hypersurface $\sigma$ as
\begin{eqnarray}
  \Pi _\psi \bxk 
          \!\!&=&\!\!n_\mu \bxk{{\partial ^R{\cal L}\lxr } \over {\partial \left[ {\partial _\mu \psi \bxk} \right]}}
          =in_\mu \bxk \bar \psi \bxk \gamma ^\mu ,\label{eq:MOM1}\\
  \Pi _W^{i\nu } \bxk
          \!\!&=&\!\!n_\mu \bxk{{\partial ^R{\cal L}\lxr } \over {\partial \left[ {\partial _\mu W_\nu ^i \bxk} \right]}}
          =-n^{\mu}\bxk\partial_{\mu} W^{i\nu } \bxk
           -g f^{ijk} ({n^\mu\bxk W_\mu ^j \bxk})W^{k\nu } \bxk,\label{eq:MOM2}\\
  \Pi _c^a \bxk
          \!\!&=&\!\!n_\mu \bxk{{\partial ^R{\cal L}\lxr } \over {\partial \left[ {\partial _\mu c^a \bxk} \right]}}
          =n^{\mu}\bxk\partial_{\mu} \bar c^a \bxk,\label{eq:MOM3}\\
  \Pi _{\bar c}^a \bxk
          \!\!&=&\!\!n_\mu \bxk{{\partial ^R{\cal L}\lxr } \over {\partial \left[ {\partial _\mu \bar c^a \bxk} \right]}}
          =-n^{\mu}\bxk\partial_{\mu} c^a \bxk-g f^{acb}\left( {n^\mu \bxk W_\mu ^c \bxk} \right)c^b \bxk, \label{eq:MOM4}
\end{eqnarray}
where $n^{\mu}\bxk$ is a unit normal $(n^\mu \bxk n_\mu \bxk = 1)$ at $x^{\mu}$
to the surface passing through $x^{\mu}$. The superscript ^^ ^^ $\ R$ " means that we have adopted
the right-differentiation convention.
If we make the space-like hypersurface to be global flat surface, 
the conjugate momenta (\ref{eq:MOM1})--(\ref{eq:MOM4})
reduce to the ordinary conjugate momenta.
Thus our conjugate momenta are natural extension of the conventional ones defined at a particular time.

The total QCD Hamiltonian density is given
by the Legendre transformation on $\sigma$:
\begin{eqnarray}
{\cal H}\lxr &=&
   \Pi _\psi \bxk\partial _n\psi \bxk +\Pi _W^{i\nu } \bxk\partial _nW_\nu ^i \bxk +\Pi _c^a \bxk\partial _nc^a \bxk
   +\Pi _{\bar c}^a \bxk\partial _n\bar c^a \bxk-{\cal L}\lxr  \nonumber\\
 &=&
  {\cal H}_{\rm 0}\lxr + {\cal H}_{{\rm int}}\lxr, \label{eq:totham}
\end{eqnarray}
with
\begin{eqnarray*}
{\cal H}_{\rm 0}\lxr =
  &-&\Pi _\psi \bxk\left( {n\cdot \gamma } \right)\gamma _\mu \partial _t^\mu \psi\bxk
  -i\Pi _\psi\bxk \left( {n\cdot \gamma } \right)\gamma ^\nu \psi\bxk \\ 
  &-&{1 \over 2}{\Pi _W}_\mu ^i \bxk {\Pi _W}^{i\mu } \bxk
  +{1 \over 2}\left( {\partial _{t\mu }W^{i\nu } \bxk} \right)\left( {\partial _t^\mu W_\nu ^i \bxk} \right) \\
  &+&\Pi _{\bar c}^a \bxk\Pi _c^a \bxk-\left( {\partial _{t\mu }\bar c^a \bxk} \right)\left( {\partial _t^\mu \bar c^a \bxk} \right),
\end{eqnarray*}
\begin{eqnarray*}
{\cal H}_{\rm int}\lxr =&&
              ig\Pi _\psi \bxk( {n\cdot \gamma } )\gamma ^\mu \lambda ^i\psi\bxk W_\mu ^i\bxk \\
          &-&g\left[ { f^{ilm}( {n\cdot W^l} )\Pi _{W\nu }^i \bxk W^{m\nu }\bxk
             -{1 \over 2} f^{ijk}W_\mu ^j \bxk W_\nu ^k \bxk\left( {\partial _t^\nu W^{i\nu }\bxk
             -\partial _t^\nu W^{i\mu } \bxk} \right) }\right] \\
          &+&{{g^2} \over 2} f^{ijk}
                   f^{ilm}\left[ {( {n\cdot W^k} )( {n\cdot W^l} )W_\mu^j \bxk W^{m\mu }\bxk
                  +{1 \over 2}W_\mu ^j \bxk W_\nu ^k \bxk W^{l\mu } \bxk W^{m\nu }\bxk} \right] \\
          &-&g f^{acb}\Pi _c^a\bxk \left( {n\cdot W^c} \right)c^b\bxk
                -g f^{acb}\left( {\partial _{t\nu }\bar c^a \bxk} \right)W^{c\nu }\bxk c^b \bxk,
\end{eqnarray*}
where
$$
\partial_{\mu} = n_{\mu}\bxk \partial_{n} + {\partial_{t}}_{\mu},
$$
$$
\partial_{n}  = n_{\mu}\bxk \partial^{\mu},\quad
{\partial_{t}}_{\mu} = (g_{\mu \nu} - n_{\mu}\bxk n_{\nu}\bxk) \partial^{\nu}.
$$
The differential operators $\partial_{n}$ and ${\partial_{t}}_{\mu}$ are the directional derivatives 
in the direction of normal and tangent to $\sigma$, respectively.
With the help of the total canonical energy-momentum tensor $T_{\mu \nu} \lxr$, 
we can express (\ref{eq:totham})
in the form
$${\cal H}\lxr = n^\mu \bxk n^\nu \bxk T_{\mu \nu} \lxr.$$ 
Thus we can identify ${\cal H}_{0}\lxr$ as the kinetic term of the Hamiltonian density, and
${\cal H}_{\rm int}\lxr$ as the interaction term of the Hamiltonian density.
The surface integral of ${\cal H}$ on $\sigma$
should be regarded as the total QCD Hamiltonian:
$$
H  =\int_{\sigma} d\Sigma^{\mu}\bxk n^{\nu}\bxk T_{\mu\nu}\lxr = \int_{\sigma} d\Sigma {\cal H},
$$
where $d\Sigma^{\mu} \bxk = n^{\mu}\bxk d\Sigma$.
We get Hamilton's equations by taking the variation of $H$ on $\sigma$:
\begin{equation}
%
%
{{\partial ^R{\cal H}} \over {\partial \Pi _\psi }}
-\partial _{t\mu }{{\partial ^R{\cal H}} \over {\partial \left[ {\partial _{t\mu }\Pi _\psi } \right]}}
+\partial _n\psi =0,\qquad {{\partial ^R{\cal H}} \over {\partial \psi }}
-\partial _{t\mu }{{\partial ^R{\cal H}} \over {\partial \left[ {\partial _{t\mu }\psi } \right]}}
+\partial _n\Pi _\psi =0, \label{eq:HA1}
\end{equation}
%
%
\begin{equation}
{{\partial ^R {\cal H}} \over {\partial W_\lambda ^i}}
  -\partial _{t\mu }{{\partial ^R {\cal H}} \over {\partial \left[ {\partial _{t\mu }W_\lambda ^i} \right]}}
  +\partial _n\Pi _W^{i\lambda }=0,\  {{\partial ^R {\cal H}} \over {\partial \Pi _{W\lambda }^i}}
  -\partial _{t\mu }{{\partial ^R {\cal H}} \over {\partial \left[ {\partial _{t\mu }\Pi _{W\lambda }^i} \right]}}
  -\partial _nW^{i\lambda }=0, \label{eq:HA2}
\end{equation}
%
%
\begin{equation}   
{{\partial ^R {\cal H}} \over {\partial \Pi _c^a}}
  -\partial _{t\mu }{{\partial ^R {\cal H}} \over {\partial \left[ {\partial _{t\mu }\Pi _c^a} \right]}}
  +\partial _nc^a=0,\qquad {{\partial ^R {\cal H}} \over {\partial c^a}}
  -\partial _{t\mu }{{\partial ^R {\cal H}} \over {\partial \left[ {\partial _{t\mu }c^a} \right]}}
  +\partial _n\Pi _c^a=0, \label{eq:HA3}
\end{equation}
%
%
\begin{equation}  
{{\partial ^R {\cal H}} \over {\partial \Pi _{\bar c}^a}}
  -\partial _{t\mu }{{\partial ^R {\cal H}} \over {\partial \left[ {\partial _{t\mu }\Pi _{\bar c}^a} \right]}}
  +\partial _n\bar c^a=0,\qquad {{\partial ^R {\cal H}} \over {\partial \bar c^a}}
  -\partial _{t\mu }{{\partial ^R {\cal H}} \over {\partial \left[ {\partial _{t\mu }\bar c^a} \right]}}
  +\partial _n\Pi _{\bar c}^a=0. \label{eq:HA4}
\end{equation}
They give field equations (\ref{eq:EL1})--(\ref{eq:EL4})
(excepting a half set of equations for gluons leading to a trivial identity $0=0$), 
so Hamilton's equations (\ref{eq:HA1})--(\ref{eq:HA4}) are equivalent to the Euler-Lagrange equations.

Hamilton's equations (\ref{eq:HA1})--(\ref{eq:HA4}) and the Klein-Gordon equation associated with
the invariant delta function ($(\dAlemb +m^2)\Delta(x)=0$ ; $\dAlemb D(x)=0$)
can be combined to give the following surface integrals
\begin{eqnarray}
  \psi (x;\sigma )&=&-\int_\sigma  {d\Sigma (i\gamma \cdot \partial +m)\Delta (x-x'')\gamma _0\Pi _\psi ^\dagger (x'')},\label{eq:s1}\\
  \bar \psi (x;\sigma )&=&-\int_\sigma  {d\Sigma \Pi _\psi (x'')}(-i\gamma \cdot \partial +m)\Delta (x-x''),\label{eq:s2}\\
  W^{i\lambda }(x;\sigma )&=&
   \int_\sigma  {d\Sigma [\partial ''_nD(x-x'')\cdot W^{i\lambda }(x'')+D(x-x'') \Pi_W^{i\lambda }(x'')}],\label{eq:s3}\\
  c^a(x;\sigma )&=&\int_\sigma  {d\Sigma [\partial ''_nD(x-x'')\cdot c^a(x'')+D(x-x'')\Pi _{\bar c}^a(x'')]},\label{eq:s4}\\
  \bar c^a(x;\sigma )&=&\int_\sigma  {d\Sigma [\partial ''_nD(x-x'')\cdot \bar c^a(x'')-D(x-x'')\Pi _c^a(x'')]}.\label{eq:s5}
\end{eqnarray}
Note that the world-point $x$ is not necessarily on the surface $\sigma$, 
and space-time coordinates on $\sigma$ are represented by $x''$.
We find that the integrals (\ref{eq:s1})--(\ref{eq:s5}) satisfy free field equations:
\begin{eqnarray}
(i \gamma \cdot \partial - m) \ \psi (x;\sigma )&=&0, \nonumber\\
\dAlemb\ W^{i\lambda }(x;\sigma )&=&0, \nonumber\\
\dAlemb\ c^a (x;\sigma)&=&0, \nonumber\\
\dAlemb\ \bar c^a(x;\sigma )&=&0. \nonumber
\end{eqnarray}
Thus the surface integrals (\ref{eq:s1})--(\ref{eq:s5}) are equivalent to asymptotic fields.

The bubble differentiation ~\cite{Tom,KTT,Schwinger}
( a differentiation with respect to $\sigma$ )
of the integrals (\ref{eq:s1})--(\ref{eq:s5}) around a world-point $x'$ reads
\begin{eqnarray}
 {{\delta \psi (x;\sigma )} \over {\delta \sigma \bxpk}}&=&
 -g(i\gamma \cdot \partial +m)\Delta ({\eightrm x-x'} )\gamma ^\mu \lambda ^i\psi \bxpk W_\mu ^i\bxpk ,\label{eq:bd1}\\
 {{\delta W^{i\lambda }(x;\sigma )} \over {\delta \sigma \bxpk }}&=&
 -\partial '^\mu D({\eightrm x-x'} )\cdot g f^{ijk}W_\mu ^j\bxpk W^{k\lambda }\bxpk 
 +\!D({\eightrm x-x'} )\!\cdot g f^{ijk}G^{j\lambda \nu }\bxpk W_\nu ^k\bxpk \nonumber\\
 & &+\!D({\eightrm x-x'} )\cdot g f^{icb}[\partial '^\lambda \bar c^c\bxpk ]c^b\bxpk 
 -D({\eightrm x-x'} )\cdot ig\bar \psi \bxpk \gamma ^\lambda \lambda ^i\psi \bxpk ,\label{eq:bd2}\\
  {{\delta c^a (x;\sigma )} \over {\delta \sigma \bxpk }}&=&
 -g f^{acb}\partial '^\mu D({\eightrm x-x'} )\cdot W_\mu ^c\bxpk c^b\bxpk ,\label{eq:bd3}\\
  {{\delta \bar c^a (x;\sigma )} \over {\delta \sigma \bxpk }}&=&
  g f^{acb}D({\eightrm x-x'} )W_\mu ^c\bxpk \partial '^\mu \bar c^b\bxpk . \label{eq:bd4}
\end{eqnarray}
The right-hand side of equations (\ref{eq:bd1})--(\ref{eq:bd4}) includes only
the interaction term, so the integrals $(\ref{eq:s1})$--$(\ref{eq:s5})$ are not conserved
because of the interaction between fields. 
However, equations (\ref{eq:bd1})--(\ref{eq:bd4})
are invariant under the canonical transformations. The invariance
is easily verified to represent equations $(\ref{eq:bd1})$--$(\ref{eq:bd4})$
in terms of Hamilton's equations and the Klein-Gordon equation of the invariant delta function.

If we impose the Lorentz condition
$\partial''_\rho W^{i\rho} ({\eightrm x''})=0$ on $\sigma$ as a boundary condition,
the supplementary condition of the $W^{i\rho }(x;\sigma )$ becomes
\begin{eqnarray}
\partial_{\rho} W^{i\rho }(x;\sigma ) - \int_\sigma  
d\Sigma _\rho \bxdpk  [ 
  &&\partial ''^\mu D({\eightrm x-x''} )\cdot g f^{ijk}W_\mu ^j\bxdpk W^{k\rho }\bxdpk \nonumber\\
  &&{ }{ }-D({\eightrm x-x''} )\cdot g f^{ijk}G^{j\rho \nu }\bxdpk W_\nu ^k\bxdpk  \nonumber\\
  &&{ }{ }-D({\eightrm x-x''} )\cdot g f^{icb}[\partial '^\lambda \bar c^c\bxdpk ]c^b\bxdpk \nonumber\\
  &&{ }{ }+D({\eightrm x-x''} )\cdot ig\bar \psi \bxdpk \gamma ^\rho \lambda ^i\psi \bxdpk 
  ]
  \approx 0. \label{eq:supp1}
\end{eqnarray}
The weak equality symbol $\approx$ is to emphasize that the left-hand side is numerically
restricted to be zero but does not identically vanish in the phase space.

Now, we define a Poisson bracket on the space-like hypersurface $\sigma$. 
Any two Lorentz covariant quantities $F[\sigma]$ and $G[\sigma]$ 
being the function of canonical field variables and momenta on $\sigma$ have a Poisson bracket 
which we shall denote by $\left[ {F[\sigma],G[\sigma]} \right]_c$, defined by
\begin{equation}
\left[ {F[\sigma],G[\sigma]} \right]_c\!\!
=\int_\sigma  \!\!\!{d\Sigma}\!\left( {{{\tilde \delta F[\sigma]} \over {\tilde \delta \phi _A \bzk}}
{{\tilde \delta G[\sigma]} \over {\tilde \delta \Pi _A \bzk}}
-(-1)^{\left| A \right|}{{\tilde \delta F[\sigma]} \over {\tilde \delta \Pi _A \bzk}}
{{\tilde \delta G[\sigma]} \over {\tilde \delta \phi _A \bzk}}} \right). \label{eq:pb}
\end{equation}
The ${\left| A \right|}$ is the number of factor associated with the classical field $\phi_A$:
take 0 for Grassmann even function, and 1 for Grassmann odd function.
If $F[\sigma]$ is given by
\begin{eqnarray*}
F[\sigma]&=&\int_{\sigma} {d\Sigma ^\nu \bxk}\ f_\nu 
(\phi _A \bxk,\partial _{\mu t}\phi _A(x),\Pi _A \bxk,\partial _{\mu t}\Pi _A \bxk)\nonumber\\
                        &=&\int_{\sigma} {d\Sigma \ {\cal F}
(\phi _A,\partial _{\mu t}\phi _A,\Pi _A,\partial _{\mu t}}\Pi _A)
\end{eqnarray*}
with $d\Sigma ^\nu f_{\nu}=d\Sigma (n\!\cdot\! f) \equiv d\Sigma {\cal F},$
the functional derivative on $\sigma$ reads
$$
{{\tilde \delta F[\sigma ]} \over {\tilde \delta \phi _A \bzk}}
={{\partial ^R{\cal F}} \over {\partial \phi _A \bzk}}-\partial _{t\mu} 
{{\partial ^R{\cal F}} \over {\partial [\partial _{t\mu} \phi _A \bzk]}},
$$
$$
{{\tilde \delta F[\sigma ]} \over {\tilde \delta \Pi _A \bzk}}
={{\partial ^R{\cal F}} \over {\partial \Pi _A \bzk}}-\partial _{t\mu} 
{{\partial ^R{\cal F}} \over {\partial [\partial _{t\mu} \Pi _A \bzk]}}.$$

Our Poisson bracket (\ref{eq:pb}) is a generalization of the ordinary equal-time Poisson bracket
to be Lorentz covariant. All algebraic relations of the ordinary Poisson bracket hold for
the Poisson bracket on $\sigma$. Here we will
call the Poisson bracket on $\sigma$ the four-dimensional Poisson bracket.

Substituting $(\ref{eq:s1})$--$(\ref{eq:s4})$ into $(\ref{eq:pb})$, we get
the four-dimensional Poisson bracket relations of $\psi(x;\sigma), W_{\mu}^i(x;\sigma), c^a (x;\sigma)$,
and $\bar c^a(x;\sigma)$:
\begin{eqnarray}
   \left[ {\psi _\alpha (x;\sigma ),\bar \psi _\beta (y;\sigma )} \right]_c
          &=&(i\gamma \cdot \partial +m)_{\alpha \beta }\Delta ({\eightrm x-y} ),\\
   \left[ {W_\mu ^i(x;\sigma ),W_\nu ^j(y;\sigma )} \right]_c
          &=&-\delta ^{ij}g_{\mu \nu }D({\eightrm x-y} ),\\
   \left[ {c^a(x;\sigma ),\bar c^b(y;\sigma )} \right]_c
          &=&\delta ^{ab}D({\eightrm x-y} ).
\end{eqnarray}
If there are no interaction, $\psi(x;\sigma), W_{\mu}^i(x;\sigma), c^a (x;\sigma)$,
and $\bar c^a(x;\sigma)$ are conserved and do not depend on $\sigma$.
It follows that
\begin{eqnarray}
   \left[ {\psi _\alpha (x),\bar \psi _\beta (y)} \right]_c
          &=&(i\gamma \cdot \partial +m)_{\alpha \beta }\Delta ({\eightrm x-y} ),\nonumber\\
   \left[ {W_\mu ^i(x),W_\nu ^j(y)} \right]_c
          &=&-\delta ^{ij}g_{\mu \nu }D({\eightrm x-y} ),\nonumber\\
   \left[ {c^a(x),\bar c^b(y)} \right]_c
          &=&\delta ^{ab}D({\eightrm x-y} ). \nonumber
\end{eqnarray}
These four-dimensional Poisson bracket relations correspond to the four-dimensional commutation relations 
for free fields in QCD.

Equations $(\ref{eq:bd1})$--$(\ref{eq:bd4})$ are also expressible in terms of
the four-dimensional Poisson bracket:
\begin{eqnarray}
  {{\delta \psi (x;\sigma )} \over {\delta \sigma \bxpk }}&=&
      -\left[ {\psi (x;\sigma ),{\cal H}_{\rm int}\lxpr} \right]_c,\label{eq:YF1}\\
  {{\delta W^{i\lambda }(x;\sigma )} \over {\delta \sigma \bxpk }}&=&
      \left[ {W^{i\lambda }(x;\sigma ),{\cal H}_{\rm int}\lxpr} \right]_c,\label{eq:YF2}\\
  {{\delta c^a(x;\sigma )} \over {\delta \sigma \bxpk }}&=&
     -\left[ {c^a(x;\sigma ),{\cal H}_{\rm int}\lxpr} \right]_c,\label{eq:YF3}\\
  {{\delta \bar c^a(x;\sigma )} \over {\delta \sigma \bxpk }}&=&
     -\left[ {\bar c^a(x;\sigma ),{\cal H}_{\rm int}\lxpr} \right]_c.\label{eq:YF4}
\end{eqnarray}
These equations show that the displacement of
$\psi (x;\sigma), W_\mu^i(x;\sigma), c^a (x;\sigma)$ and $\bar c^a (x;\sigma)$
under the deformation of $\sigma$ is equal to that of the surface integrals under
the canonical transformation generated by the interaction Hamiltonian density ${\cal H}_{\rm int}$
at a world-point $x'$ on $\sigma$.

To pass over to QCD in the Heisenberg picture, we shall make all
canonical variables into operators and convert all the four-dimensional Poisson bracket relations
to four-dimensional commutation relations:
$$
{\hbox{(four-dimensional Poisson bracket)}} \rightarrow {(i\hbar)}^{-1}{\hbox{(four-dimensional commutator)}}.
$$
In this way we can pass over to QCD without spoiling the Lorentz covariance.
A remarkable feature of
our prescription in quantization is that equations ({\ref{eq:YF1}})--({\ref{eq:YF4}}) are 
converted to the Yang--Feldman equations~\cite{YF}. 
According to the procedure given by Yang and Feldman, 
the S-matrix of the Heisenberg picture in QCD must be identified as
$$
{\bf S}=\cdots 
\left[ {1-i\int_{\sigma _{-1}}^{\sigma _0} {{\cal\widehat H}_{\rm int}\lxpr\ d^4x'}} \right]
\left[ {1-i\int_{\sigma _0}^{\sigma _1} {{\cal\widehat H}_{\rm int}\lxpr\ d^4x'}} \right]\cdots, 
$$
where ${\cal\widehat H}_{\rm int}\lxpr$ is the operator corresponding to 
the interaction Hamiltonian density ${\cal H}_{\rm int}\lxpr$.

The supplementary condition (\ref{eq:supp1}) will be converted to
\begin{eqnarray}
\Bigl[ \partial_{\rho} W^{i\rho }(x;\sigma ) -\int_\sigma  
  d\Sigma _\rho \bxdpk  \bigl[ 
  &&\partial ''^\mu D({\eightrm x-x''} )\cdot g f^{ijk}W_\mu ^j\bxdpk W^{k\rho }\bxdpk \nonumber\\
  &&{ }{ }-D({\eightrm x-x''} )\cdot g f^{ijk}G^{j\rho \nu }\bxdpk W_\nu ^k\bxdpk  \nonumber\\
  &&{ }{ }-D({\eightrm x-x''} )\cdot g f^{icb}[\partial '^\rho \bar c^c\bxdpk ]c^b\bxdpk \nonumber\\
  &&{ }{ }+D({\eightrm x-x''} )\cdot ig\bar \psi \bxdpk \gamma ^\rho \lambda ^i\psi \bxdpk 
  \bigr] \Bigr]^{(+)}\vert \Psi >
  = 0 \label{eq:spc2},
\end{eqnarray}
where the symbol $(+)$ means the positive-frequency part, and $\vert \Psi>$ is a state vector
in the Heisenberg picture.

\vspace {0.5 cm}
I would like to thank Professor M. J. Hayashi for careful reading
of manuscript.



\begin{references}

\bibitem{Tom}S. Tomonaga, Prog. Theor. Phys. {\bf 1}, 27 (1946).
  
\bibitem{KTT}Z. Koba, T. Tati, and S. Tomonaga, Prog. Theor. Phys. {\bf2}, 101 (1947).
  
\bibitem{Schwinger}J. Schwinger, Phys. Rev. {\bf 74}, 1439 (1948).
 
\bibitem{YF}C. N. Yang and D. Feldman, Phys. Rev. {\bf 79}, 972 (1950).
  
\end{references}
\end{document}